\documentclass[twocolumn,aps,prd,epsf]{revtex4}
\usepackage{epsfig}

\newcommand{\be}{\begin{equation}}
\newcommand{\ee}{\end{equation}}
\newcommand{\bea}{\begin{eqnarray}}
\newcommand{\eea}{\end{eqnarray}}

\begin{document}
\title{Glueball and hybrid mass and decay with string tension below Casimir scaling}
\author{
Elsa Abreu }
\email{elsabreu@netcabo.pt}
\affiliation{ Dep. F\'{\i}sica and CFIF, Instituto Superior T\'ecnico,
Av. Rovisco Pais, 1049-001 Lisboa, Portugal}
\author{
Pedro Bicudo }
\email{bicudo@ist.utl.pt}
\affiliation{ Dep. F\'{\i}sica and CFIF, Instituto Superior T\'ecnico,
Av. Rovisco Pais, 1049-001 Lisboa, Portugal}
\email{bicudo@ist.utl.pt}

\begin{abstract}
Lattice computations with excited SU(3) representations suggest that the 
confining gluon-gluon interaction complies with the Casimir scaling. 
The constituent gluon models have also been assuming the Casimir scaling. 
Nevertheless, inspired in type-II superconductors, we explore a 
new scenario for the gluon-gluon interaction where the
adjoint string is replaced by a pair of fundamental strings,
resulting in a factor of 2, smaller than 9/4.
To test our proposal we construct a simple constituent gluon model,
extrapolated from the funnel potential for quarkonium,
and apply it to compute the wave-function of glueballs and of hybrid gluelumps. 
From the decay widths of quarkonium, we also extrapolate the decay
widths of the glueballs. Our predictions apply to
charmonia, lightonia, glueballs and hybrid gluelumps with large angular momentum. 
\end{abstract}

\maketitle
\section{Introduction}

The glueballs are expected from QCD, and indeed they are
observed in pure gauge lattice QCD simulations. 
The first experimental evidence for glueballs is an indirect 
one, the pomeron which explains the high energy scattering
of hadrons,
\cite{Collins,Donnachie_book,Simonov,weall}.
Systematic comparisons of the glueball masses computed in
Lattice QCD with the pomeron trajectory, initiated by
Llanes-Estrada, Cotanch, Bicudo, Ribeiro and Szczepaniak, 
confirm that the lattice QCD glueballs comply with the pomeron
\cite{weall,Meyer,Morningstar,Liu}.
Direct experimental evidence for glueballs is still controversial
\cite{Bugg},
nevertheless in few years two major
collaborations will update the experimental search for glueballs, 
PANDA at GSI and GLUEX at JLAB. Different theoretical 
approaches to glueballs are trying to match the experimental
effort. 

\begin{figure}[t]
\begin{picture}(220,300)(0,0)
\put(0,-40){\epsfig{file=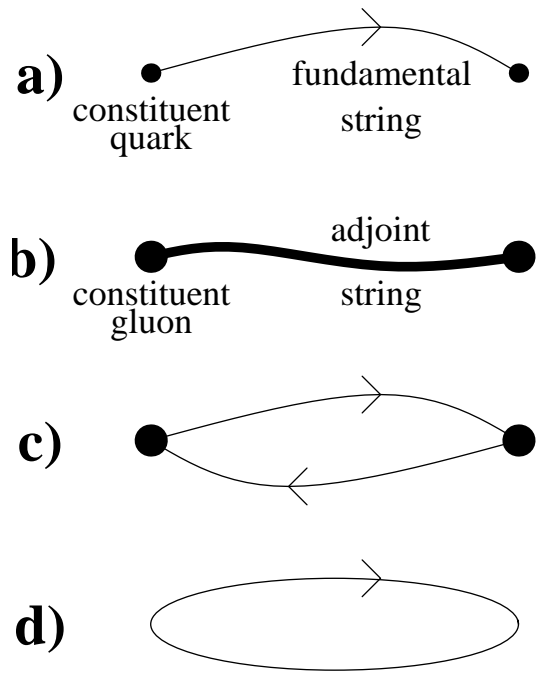,width=8.6cm}}
\end{picture}
\caption{Basic models b) c) and d) for glueballs, extended
from the string model a) for quark-antiquark confinement in a meson} 
\label{3 models}
\end{figure}

Here we aim at a simple and plausible model for glueballs
and for other hadrons, with large angular excitations. 
Many of these hadrons remain to be observed, see Table \ref{widths}.
We first review the possible glueball models, extending the
string mechanism for quark-antiquark confinement. The quark
model describes phenomenology, simply assuming constituent quarks, 
a short range coulomb interaction and a long range confining string
interaction. This model is depicted in Fig. \ref{3 models} $a)$.
Nambu and Jona-Lasinio 
\cite{Nambu}
showed that the massless fermions can 
have a mass generated by the spontaneous breaking of chiral symmetry,
with the Schwinger-Dyson technique, equivalent to a second-quantized
canonical transformation
\cite{Yaouanc,Adler,Bicudo_thesis}.
Indeed this has been successfully applied to quark models
\cite{Yaouanc,Adler,Bicudo_thesis},
suggesting that quark models may comply both with the PCAC 
theorems and with the success of quark model spectroscopy
\cite{Scadron,weall2,Bicudo_PCAC,Llanes_chilag}.
Moreover the quark-antiquark confining string, consisting of a 
color-electric flux tube was suggested by Nielsen and Olesen
\cite{Nielsen}. 
Nielsen and Olesen conjectured that the magnetic flux tube
vortex of type-II superconductors could be extended to
color-electric flux tube strings in QCD.

%
%

Inspired in the constituent string-confined quark model $a)$, three
models for glueballs are plausible. They are respectively depicted
in Fig. \ref{3 models}. 
In model $b)$, 
a constituent gluon pair is confined with an adjoint string,
in model $c)$, 
a constituent gluon pair is confined with a pair of fundamental strings
and in model $d)$,
a closed fundamental string loop is rotating , without any constituent
gluon at all.
Models $b)$
\cite{Simonov,weall,Brisudova,Szczepaniak,Brau}, 
with a pair of constituent gluons confined
by an open adjoint string, and $d)$
\cite{Niemi,Meyer_thesis}, 
with a closed fundamental string and no constituent gluon, 
and also hybrid models
\cite{Kalashnikova,Simonov_hybrid},
have been studied in the literature.
The model $c)$ is explored in this paper.

An important ingredient of models $b)$ and $c)$ is the massive constituent
gluon.  There is evidence for a massive-like dispersion relation
for the gluon, with a mass ranging from 700 MeV to 1000 MeV,
both from laticce QCD
\cite{Bernard,Stella,Bonnet,Oliveira}
and from Schwinger-Dyson equations
\cite{Alkofer}.
This suggests that the Anderson-Higgs mechanism is
providing a consistent framework for the mass generation
of gauge bosons. For instance in the Meissner effect of
superconductors, the photon has a mass.
It is plausible that simple non-relativistic constituent
gluon models may indeed describe glueballs. 

%
%
\begin{table}[t]
\caption{ \label{widths}
Experimental (Review of Particle Physics) available masses M and widths
$\Gamma$, for charmonia, lightonia, $K^*$, glueballs and gluelumps, 
with maximal $s$ and $j=l+s$.
}
\begin{ruledtabular}
\begin{tabular}{|l|cccccc|} 
states / $J^{PC}$ & $0^{++}$ & $1^{--}$ & $2^{++}$ & $3^{--}$ & $4^{++}$ & $5^{--}$  \\
 \hline
Charmonium, M  			&   & 3096 & 3556 & - & - & - \\
Charmonium, $\Gamma$	&   & 0.1 & 2.1 & - & - & - \\
lightonium I=0, M  		&   & 783 & 1275 & 1667 & 2034 & - \\
lightonium I=0, $\Gamma$&   & 8.5 & 185 & 168 & 222 & - \\ 
lightonium I=1, M  		&   & 775 & 1318 & 1689 & 2010 & - \\
lightonium I=1, $\Gamma$&   & 150 & 107 & 161 & 353 & - \\ 
lightonium K* , M  				&   & 892 & 1430 & 1776 & 2045 & - \\
lightonium K* , $\Gamma$			&   & 51  & 105 & 159 & 198 & - \\ 
Glueball, M  			&   &   & - &   & - & \\
Glueball, $\Gamma$  	&   &   & - &   & - & \\
Gluelump, M  			&   &   & - & - & - & - \\
Gluelump, $\Gamma$  	&   &   & - & - & - & - \\
\end{tabular}
\end{ruledtabular}
\end{table}

Another ingredient, necessary for model $b)$ only, is the adjoint string.
The adjoint string is a natural extension of the fundamental string
assumed in models $a), \, c), \, d)$. It consists of an excited string
with a colour-electric flux larger than the one of the fundamental string.
Two possible contributions for colour-colour interaction are
the shorter range one gluon exchange interaction, 
and the longer range confining string interaction, proportional to the 
colour-electromagnetic energy density.
Both are proportional to $\lambda \cdot \lambda$, one of the 
two QCD Casimir invariants. Moreover Bali 
\cite{Debbio,Bali}
showed, in pure-gauge 
lattice QCD, extending the lattice gauge field links from the fundamental representation
to several representations of SU(3), that the colour-colour interactions
comply with the Casimir scaling, they are all proportional 
to $\lambda \cdot \lambda$, except for smaller than 5\% numerical errors.

Nevertheless we propose here a new
model $c)$ with the same pair of constituent gluons of $b)$, 
but with a gluon-gluon potential lower
\cite{Bicudo_PANDA}
than the Casimir scaling of model $c)$.
Importantly, model $c)$ is not excluded by the lattice QCD computations
published so far. The Casimir scaling for the octet-octet interaction was observed in 
lattice discretizations of QCD where the fundamental links are replaced by adjoint links, 
$8 \times 8$ matrices of the adjoint representation of $SU(3)$. Therefore 
these lattice computations are not adequate to simulate our model $c)$ which 
assumes fundamental strings only.  

Like the seminal paper of Nielsen and Olesen
\cite{Nielsen}, 
model $c)$ is inspired in type-II 
superconductors
\cite{Gennes}. 
In type-II superconductors the magnetic flux is 
allowed to penetrate, and it is localized in vortices,
with flux 
\be
\phi_n= n { h c \over 2 e} \ ,
\ee
where the number $n$ quantifies the magnetic flux. 
Rather than having a single vortex with a high flux, it is energetically 
favourable to have several elementary vortices, say a square lattice of 
vortices, each with the minimal flux $\phi_1$. This occurs 
because the energy of each vortex is proportional to the square of the 
magnetic flux. For instance a vortex with $n=2$ does cost the double of 
the energy of two elementary vortices with $n=1$, while
both scenarios have the same flux $\phi_2=2 \phi_1$.
Back to our constituent gluon potential, it is energetically favourable
to have two fundamental strings, with scaling factor of $2$,
than one adjoint (excited) string, with a scaling factor of $9/4=2.25$.

In the remaining of this paper we specialize in the proposed model $c)$, 
extending the basic constituent quark model, from the 
quark-antiquark meson, to the gluon-gluon glueball and to the 
quark-antiquark-gluon hybrid. Because we also study light mesons,
our kinetic energy is relativistic. We specialize in systems with maximal 
angular momentum, because the string length is large and the details of the
confining string become relevant. These large $J$ states also allows us to avoid 
the details of spin tensor potentials. The experimentally confirmed resonances 
only go up to $J=4$ for light mesons and $J=2$ for charmonium, see 
Table \ref{widths}, therefore most of our results are predictions. 

In Section II we discuss the parameters of our model.
The scaling of 2, the experimental meson masses, the glueball masses
an parameters obtained in the lattice and in the Schwinger-Dyson 
equations, together with the pomeron intercept determine 
all the parameters of the model. 
In Section III  we compute the masses (up to 5 GeV, close to the energy limit of 
the future experiments) and the mean radius of charmonia, lightonia, glueballs and hybrids. 
We also estimate decay constants in Section IV.
In Section V we conclude.

\section{The parameters of this constituent quark and gluon model}

\subsection{Mesons}

Our starting point to parametrize the potentials and masses is
the funnel potential for charmonium
\cite{Lucha}.
The charmonium spectrum and wavefunctions are usually determined solving the
non-relativistic radial Schr\"odinger equation for two-body systems, 
\begin{eqnarray}
&\Bigl[& M_t\, c^2 -{\hbar^2 \over 2 \mu }{d^2 \over d \, r^2 } 
+ {\hbar^2 \over 2 \mu  }{l(l+1) \over  r^2 } +V(r) 
\Bigr] \nu_{n\,l}(r) = E_{n\, l} \nu_{n\,l}(r) \ ,
\nonumber \\
&&\nu_{n\,l}(r)= r R_{n\,l}(r)   \ , \ M_t=m1+m2 \ , \ \mu={m_1 \, m_2 \over m_1+m_2} \ .
\label{non rel}
\end{eqnarray}
However we are also interested high angular excitations, and therefore it is more convenient 
to upgrade the kinetic energy to the relativistic one. In the center of mass frame 
this amounts to replace in eq. (\ref{non rel}),
\begin{equation}
M_t \, c^2 + {\mathbf p^2 \over 2 \mu} \rightarrow 
\sqrt{m_1^2 \, c^4 + \mathbf p^2\, c^2} + \sqrt{m_2^2 \, c^4 + \mathbf p^2\, c^2} \ .
\label{relativistic}
\end{equation}
Again we can perform the standard angular separation of the Schr\"odinger equation, where
the squared momentum is separated in a simple second derivative and in a centrifugal barrier, 
\begin{equation}
\mathbf p^2 {1 \over r} \nu_{n\,l}(r) Y_l^m(\hat r)= {\hbar^2 \over r} \left[-{d^2 \over d \, r^2 } 
+ {l(l+1) \over  r^2 } \right]\nu_{n\,l}(r) Y_l^m(\hat r) \ ,
\label{mom angular}
\end{equation}
thus the equation remains a radial equation, the spherical harmonics can be factorized.
The arguments of the square roots in eq. (\ref{relativistic}) are positive definite and therefore
there is no technical difficulty in computing the relativistic kinetic energy (\ref{relativistic}).
A possible method of finding a function $f(M)$ of a matrix $M$ uses the eigenvalues $\lambda_i$ and 
eigenvectors,
\begin{equation}
f(M)= \sum_i | i \rangle f(\lambda_i) \langle i | \ .
\label{fundiag}
\end{equation}
Thus, once the squared momentum in eq. (\ref{mom angular}) is diagonalized, either 
with spherical Bessel functions or with finite differences, the relativistic kinetic energy operator
of eq. (\ref{relativistic}),
a function of the squared momentum, can be computed with the method of eq. (\ref{fundiag}) . 
Then we finally solve the schr\"odinger equation, diagonalizing the hamiltonian. Numerically the 
relativistic code is only twice as slow as the non-relativistic one, because we perform two diagonalizations. 
Importantly, the relativistic hamiltonian is adequate to stydy lighter systems, with light quarks or gluons.

The parameters of this model include the charm $m_1$ and anti-charm $m_2$
identical masses, and a Coulomb potential, a linear potential, and a constant potential,
\begin{equation}
V(r) = -{\alpha \over r } + \sigma r + {\cal C} \ .
\label{potential}
\end{equation}
The Coulomb potential $-{\alpha \over r }$  can be extracted from the
perturbative QCD one gluon exchange potential and from the non-pertubative
Luscher fluctuations of the string
\cite{Luscher}. 
All Coulomb contributions are summed in the parameter $\alpha$. 
The linear potential is parametrized by the string tension $\sigma$.
Both can be extracted from lattice QCD. Moreover the linear potential needs 
some Coulomb interaction to linearize the Regge trajectory in the Chew-Frautschi plot.
A constant ${\cal C}$ term in the potential is also needed to fit the spectrum.
Here the ${\cal C}$ sums the effect of spin-tensor potentials and of vacuum
effects. Moreover, in Lattice QCD the constant term in the potential is not determined,
and in a sense this constitutes a free parameter fo the funnel potential.
Here we use the parameters $m_1=m_2$, $\alpha$, $\sigma$ and obtained
by Lucha and Sch\"oberl
\cite{Lucha} 
to fit the the charmonium spectrum. Because our kinetic energy is relativistic, we
refit the constant ${\cal C}$ . Our charmonium parameters are show in Table \ref{parpot} .

Our next class of hadrons is the lightonia, mesons composed of light quarks.
In this case the masses are expected to be slightly larger than the third of the
lighter baryon masses, see for instance the modelling of baryons by Isgur and Karl
\cite{Isgur}. 
In what concerns the Coulomb and string tension they may
be slightly different than the ones of Charmonium.  The parameters
that we used to fit the experimental spectrum are again show in 
Table \ref{parpot}. While the same Charmonium string tension can be used,
a larger Coulomb potential is needed to linearize the light meson
Regge trajectory, and the constant potential is also slightly refitted. 
This is consistent with the Fermi-Breit short range Coulomb potential, 
which increases for light masses.

\subsection{Glueballs and hybrids}

Our next class of hadrons is the one of glueballs. 
The mass of the constituent gluon is determined with the
Schwinger-Dyson equation for the gluon propagator, 
\cite{Alkofer}
equivalent to the Bogoliubov-Valation mass gap equation
\cite{Cotanch}
for the gluon propagator, and with lattice QCD computations
of the gluon propagator 
\cite{Bonnet,Oliveira}.
Althought the gluon propagator is not gauge invariant, and it 
is not clear whether gauge symmetry is broken or not, there is 
evidence for a mass ranging from 700 MeV to 1 GeV in the dispersion 
relation of the gluon. 
This scale is also present in the lattice QCD
determinations of the first excitation of the quark-antiquark
or three-quark potentials
\cite{Suganuma,Juge}. 
Here we use a mass of 800 MeV,
lighter than the mass of a light quark-antiquark pair. In
this case the constituent gluon might be relatively stable 
against the decay to a quark-antiquark pair.

To address the gluon-gluon potential, we first review
the Casimir scaling.
In the fundamental representation of SU(3), acting
in the quark colour triplet, the Lie Algebra is represented by the $3\times 3$ Gell-Mann 
fundamental matrices $\lambda^a$. The product of the fundamental $\lambda^a$ is,
\begin{equation}
\lambda^a \lambda^b = {2 \over 3} \delta ^{a \, b } I + d^{a \, b \, c} \lambda^c + i f^{a \, b \, c} \lambda^c \ ,
\label{color product}
\end{equation}
where the antisymmetric structure constants $f^{a \, b \, c}$  are
antisymmetric and independent of the representation.
The constant tensors ${2 \over 3} \delta ^{a \, b }$ and $d^{a \, b \, c}$
are symmetric and depend on the representation.
Another relevant representation is the adjoint representation.
In the adjoint representation, acting in gluons,
the Lie Algebra matrices are represented by the $8\times 8$
adjoint matrices,
\begin{equation}
\left[ \lambda^a_{adjoint} \right]_{bc}= 2 i \, f^{b \, a \, c} .
\end{equation}
With eq. (\ref{color product}), and using the properties of 
the antisymmetric $f^{a \, b \, c}$ and of the symmetric 
$d^{a \, b \, c}$ we find that the Casimir invariant in  
the fundamental and adjoint representation are respectively, 
\begin{eqnarray}
\lambda^a \lambda^a &=& {16 \over 3} I \ ,
\nonumber \\
\lambda^a_{adjoint} \lambda^a_{adjoint} &=& 12 \,  I \ ,
\end{eqnarray}
and they differ by a factor of $9/4$, the Casimir scaling.
The Casimir scaling was indeed observed by Bali
\cite{Bali}
in lattice QCD. Bali showed that the energy density is proportional
to the $\lambda^a \lambda^a$.
This Casimir Scaling is the simplest possible result, and Semay
\cite{Semay}
showed that it occurs if the color-electric flux-tube thickness is 
essentially the same for all possible quantized fluxes. 

Nevertheless two fundamental strings cost less energy than one adjoint
string. The scaling of the present model is $2 < 9/4$. Because
the mass of our constituent gluon is between the mass of the
charm and light quarks, and the charmonia and lightonia used the
same string tension, for the string tension of the glueballs we double 
the parameter $\sigma$ already used for the quarks.

The parameter $\alpha$ of the glueball Coulomb potential can also be
indirectly estimated from the masses of glueballs in lattice QCD. The Lattice 
QCD result for the hyperfine splitting between the masses of the $2^{++}$ and 
$0^{++}$ is of the order of 0.7 GeV. This splitting can be compared to the
hyperfine splitting between the $1^{--}$ and $0^{-+}$ mesons,
which is of the order of 0.4 GeV for light mesons (once chiral effects 
are subtracted) and is of the order of 0.1 GeV for charmonium.
Using the one gluon exchange hyperfine splitting potential proportional to 
$  { \alpha  \, \vec \lambda_1 \cdot \lambda_2 \, \delta(r) \over m^2 }   \vec S_1 \cdot \vec S_2$,
similar to the one derived for mesons \cite{Eichten}, 
Cornwall and Soni and Kaidalov and Simonov
\cite{Cornwall,Simonov},
found an hyperfine splitting, between the $2^{++}$ and $0^{++}$ 
glueballs, of the order of 1.0 Gev.
Assuming that the hyperfine splitting potential is essentially
due to the pertubative one gluon exchange potential,
this suggests that the Coulomb potential should be suppressed by
30\%.  However, with the scaling factor of 2 , rather than the Casimir
Scaling of 2.25, we already decrease the hyperfine potential by a factor
of more than 10\%. Moreover these authors used gluon masses of the
order of 0.5 to 0.6 GeV. Our gluon masses are larger (because they are
estimated in Lattice QCD and in Schwinger Dyson truncated QCD 
equations) and this further decreases the hyperfine potential. Thus
we are confident that a glueball Coulomb potential with $2 \times$
the same $\alpha$ already used in charmonium is consistent with a
correct glueball hyperfine splitting. 

In what concerns the constant potential ${\cal C}$, it is fitted
with the intercept of the glueball trajectory, which is expected
to reproduce the pomeron.
The equation for the pomeron trajectory in the $ J,\, t=M^2$ space is
\cite{Donnachie,Pelaez},
\begin{eqnarray}
J= \alpha_p(t),
\\
\alpha_p(t) = 1.08+0.25 t \ .
\end{eqnarray}
The intercept $\alpha_p(t)$ is of the order of 1, and this explains 
the high energy hadronic cross sections. In particular we may 
ignore the small decimal digits .08 because they may be due to double
pomeron exchange. The pomeron is also expected 
to correspond to a series of glueball masses.  
Therefore our final parameter ${\cal C}$ is fitted to the intercept
$\alpha_p(0)=1$. 
Again our glueball parameters are shown in Table \ref{parpot}.


%
%
\begin{table}[t]
\caption{ \label{parpot}
 Parameters of the constituent model for glueballs, charmonium and gluelumps.
}
\begin{ruledtabular}
\begin{tabular}{|l|ccccc|} 
 parameters   & $m_1$ [GeV] & $m_2$ [GeV] & $\alpha$ & $\sigma$ [GeV$^2$] & $\cal C$ [GeV] \\
 \hline
charmonium 	& 1.64 & 1.64 & 0.27 & 0.25 & -0.69 \\
lightonium 	& 0.44 & 0.44 & 0.35 & 0.25 & -0.80 \\
glueball  	& 0.80 & 0.80 &  $2  \times$0.27  & $2 \times$0.25  & -0.52  \\
gluelump  	& 0.80 & 3.28 & $2 \times$0.27 & $2 \times$0.25  & -1.59  
\end{tabular}
\end{ruledtabular}
\end{table}

%
%
\begin{figure}[t]
\begin{picture}(300,180)(0,0)
\put(0,0){\epsfig{file=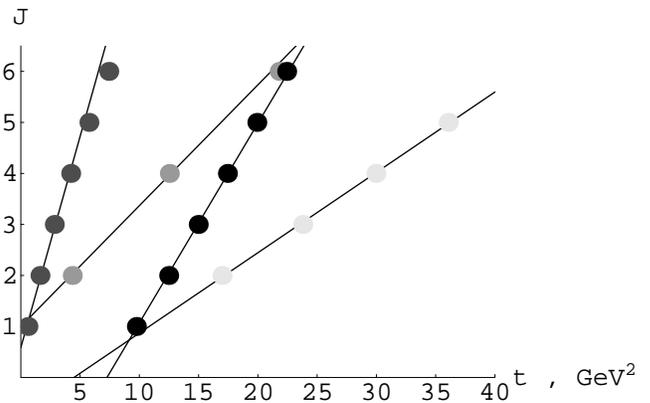,width=8.6cm}}
\end{picture}
\caption{
Chew-Frautschi plots of the leading trajectories for charmonia (black),
lightonia (dark), glueballs (grey) and hybrid gluelumps (light).
} 
\label{four trajectories}
\end{figure}

In what concerns the hybrids, it is possible that we have a gluelump.
The charm and anticharm pair may form a colour octet with a mass 
of 3.3 GeV, quite larger that the constituent gluon mass. Then the
charm-anticharm octet is essentially stopped at the centre of the mass
of the hybrid system, with the glueball orbiting it, attached by a double
fundamental string. Such a system of a gluon attached to a heavy colour
octet is called a gluelump
\cite{gluelump}. 
Theoretically, this is a very interesting object
because it is even simpler than a Glueball.
All the parameters of the gluelump potential
are the same of the glueball potential, except for the constant potential
${\cal C}$ witch may be refitted. 

The constant potential ${\cal C}$ is fitted to produce a splitting
of 1 GeV between the first gluelump state and the first $S=1$
charmonium state. This splitting has been observed in lattice
QCD for quenched quark potentials
\cite{Suganuma}. 
Since in the gluelump we ignore
the motion of the charm and the anticharm, the constant ${\cal C}$ 
not only accounts for the spin-tensor interaction, and for vacuum
effects, it also accounts for the kinetic and potential energy 
internal to the $c\bar c$ pair. Again our gluelump parameters are 
shown in Table \ref{parpot}.

\section{masses and radii of glueballs, hybrids and quarkonia with large J}

We solve the radial Schr\"odinger equation, with relativistic kinetic energies 
(\ref{relativistic}), with a finite difference method, which transforms the 
differential equation in a simple matrix eigenvalue Method. 
The parameters of our constituent model for glueballs, 
charmonium and gluelumps are show in Table \ref{parpot}.

We are particularly interested in large angular momentum states,
with maximal $S=S_1+S_2$ and with $J=L+S$. 
The maximal spin is $S=1$ for the lightonium and charmonium.
The maximal spin is $S=2$ for the glueball and for the gluelump.
We do not compute other spin-orbit combinations
or the radial excitations, because they belong to daughter trajectories,
with larger decay widths, which are harder to identify experimentally. 

The possible quantum numbers of the glueball are further constrained, 
because the gluons are bosons. Since in a colour singlet the colour
wavefunction is symmetric, and because we are only considering here the
maximal $S=2$ which is also symmetric, the angular wavefunction needs
also to be symmetric. Therefore the maximal $J$ trajectory starts
at $J=2 (L=0)$ and continues with $J=4 (L=2), \ J=6 (L=4) \cdots$

In what concerns the gluelump, the spin of the charm and anticharm pair 
is not expected to affect the gluelump mass significantly. For a maximal 
$S$ we consider that the
$c\bar c$ pair has spin 1 and that the total spin of the $c \bar c g$
system is 2. Then the leading trajectory has $J=2 (L=0), \ J=3 (L=1), 
\ J=4 (L=2) \cdots$ 

We compute the energies $E_{0\, l}$ (up to the first state above 5 GeV) and the 
corresponding wave-functions $R_{0\, l}(r)$. The energies
and the radius mean square are displayed in Tables \ref{ener} and \ref{radi}.
All our units are in powers of $GeV$, for
$\hbar=c=1$. For instance $r=10GeV^{-1}=1.97 fm$.

We also compare our glueball trajectory with the one with the 
Casimir scaling. The potential $\cal C$ is refitted in the Casimir scaling case, 
to maintain the intercept equal to 1. The new constant potential is
${\cal C}= -0.39$, and only the slope of the Chew-Frautschi trajectory changes.  
Our model {\em c)}, with a pair of fundamental strings, yields the trajectory
\begin{equation}
j=1.00 + 0.24 t \ ,
\end{equation}
in particular the slope is close to the slope of $0.25$ of the soft pomeron model
of Donnachie and Landshoff 
\cite{Donnachie}.
The model {\em b)} with Casimir scaling would produce the trajectory,
\begin{equation}
j=1.00 + 0.20 t \ .
\end{equation}
For the same gluon masses of 0.8 GeV, model {\em b)} with Casimir scaling has a smaller slope, 
smaller than the Donnachie slope.

The four different trajectories, respectively for the charmonium, lightonium,
glueball and hybrid are also displayed in Fig. \ref{four trajectories}.

%
\begin{figure}[b]
\begin{picture}(300,150)(0,0)
\put(0,0){\epsfig{file=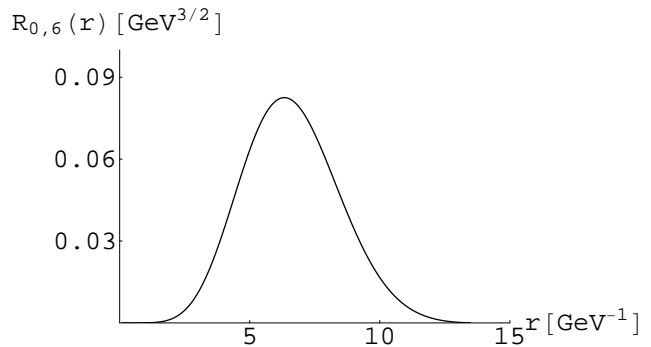,width=8.6cm}}
\end{picture}
\caption{Example of the radial part $R_{n\,l}(r)$ of the wave-function for a glueball with
a high angular excitation (l=6). 
The total wave-function is obtained multiplying the radial wave-function by the spherical harmonic
$Y_l^m(\theta,\phi)$} 
\label{Pot1}
\end{figure}

\section{Estimating decay widths of mesons and glueballs with large J}

We now estimate the decay width of the predicted hadrons, and this includes
several hadrons with large $J$ which have not yet been detected.
Let us consider for instance the decay of a glueball
\cite{Jin_decay,Cotanch_decay}.
We may assume that the decay is initiated when, 
either a gluon in the colour-electric flux tube,
or a ''massive'' constituent gluon, 
is transformed into a quark-antiquark pair with the QCD quark-gluon
coupling.

Then the glueball is transformed into an excited meson or
a hybrid. Because this is not a stable system, it also decays,
again with the same quark-antiquark pair creation. 
Essentially we have a cascade of decays, because the first decay 
products are unstable systems. This accounts for the final decay 
product of several pseudoscalar mesons 

Let us consider the decay scenario where string breaking dominates over
the direct constituent gluon decay. If the constituent gluon mass is lower 
than the constituent quark-antiquark mass, such a dominance is possible.
Here $m_g=0.8$ GeV $\le 2 \, m_q = 0.88$ GeV.
While the problem of the constituent gluon decay remains
to be solved, here we neglect this effect. Therefore our results for
the decay widths are lower bounds on the actual decay width.

In this scenario a string is cut and a $\bar q q$ pair is created. 
Each quark remains at one end of the open parts of the string.
Statistically, it is plausible that the probability to produce a $\bar q q$ pair is 
proportional to the string length. It occurs that 2x the string length in the glueball $g \, g$, 
is close to the string length in the $q \,  \bar q$ case, with the same angular momentum. 
Thus, in the glueball with a double fundamental string,
the total string length is similar to the string length of the light 
meson with the same angular momentum.
So the probability to have the breaking of a string is similar in these glueballs 
with $J^{PC}=2^{++}, \,4^{++}, \,6^{++}... $ is respectively similar to the probability
of having a string breaking in light mesons with $J^{PC}=1^{--}, \,3^{--}, \,5^{--}... $.

To compute the decay widths, one would still have to evaluate the overlaps with the 
decay products, and the available phase space. In multiple particle products,
it remains difficult, in the present state of the art of glueballs, to
study quantitatively the decays. Therefore we will essentially
focus in the relation of the string length with the decay width.
 
\begin{figure}[t]
\begin{picture}(300,180)(0,0)
\put(0,0){\epsfig{file=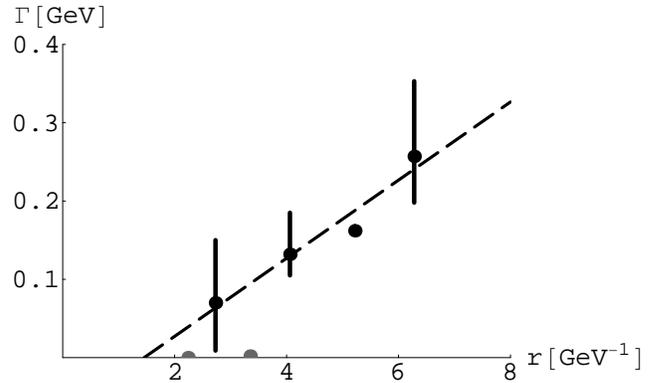,width=8.6cm}}
\end{picture}
\caption{Decay width of angular excited resonances as a function of the theoretically estimated RMS. 
The decay widths are extracted from the Review of Particle Physics
\cite{Particle Booklet}. 
The black dots are the average of the different $I=0$, $I=1$ and $K^*$ lightonium decay widths.
The vertical lines span from the maximum to the minimum of these lightonium decay widths.  
The dashed line corresponds to the best linear fit of the light data.
The gray dots represent the charmonium decay widths.
} 
\label{gamRMS}
\end{figure}

For an educated guess it is convenient to study the decays of conventional mesons.
The decay widths extracted from the Review of Particle Physics
\cite{Particle Booklet} 
are show in the Table \ref{widths}. 
The corresponding string lenghts can be approximated by the $RMS$ of Table \ref{radi}.
To show the dependence of the decay width in the string length, a 
plot is drawn in Fig. \ref{gamRMS}.  

In Fig. \ref{gamRMS} it is clear that the decay width of the light mesons is well 
fitted by a linear increase with the string length. Fitting the 
average of the different $I=0$, $I=1$ and $K^*$ lightonium decay widths we get
\begin{equation}
\Gamma = \gamma ( RMS-r_0) \pm \Delta \Gamma\ ,
\label{regression}
\end{equation}
where $\gamma =0.05 \pm 0.01$ GeV$^2$ and $r_0=1.4 \pm 0.6$ GeV$^{-1}$. 
The error in the parameters $\gamma$ and $r_0$ are statistical errors. 
The error $\Delta \Gamma =0.1 GeV$ includes the details of the decay processes.

If we extrapolate this linear growth to mesons with higher angular 
excitations, the decay width grows up to 0.68 GeV for the J$^{PC}$=15$^{--}$
meson made of light quarks, with a RMS of 15.24 GeV$^{-1}$.
The decay widths of the light mesons can also be extrapolated to glueballs.
In this case, because the glueball has a double fundamental string, for a
similar RMS, the glueball has a string lenght twice as long. Thus, for glueballs, 
the input variable $RMS$ of eq. (\ref{regression}) must be doubled. 
For instance, decay widths respectively of 0.14, 0.33 and 0.48 GeV are expected for 
the first three glueballs of our series, providing the string-breaking decay process 
is dominant. If the direct decay of constituent gluons turns out to be important, the 
glueball decay widths are larger. Our predicted widths for lightonia and for glueballs 
are shown in Table \ref{decays}. Essentially, we predict that the decay widths of both 
lightonia and glueballs with large angular excitations are of the order of 10\% of the 
respective masses.

However, in what concerns charmonium, the $D$ mesons are quite heavier than the light mesons.
The phase space of the charmonium decay products is quite different from the 
phase space of the lightonium decay products, and we should not apply this simple
linear rule, with the same parameters, to the charmonium decays and to the gluelump decays.
Moreover, only the $1^{--}$ and $2^{++}$ charmonium decay widths are known, and this
is not sufficient for a precise linear fit of the decay widths.

%
%

\section{Conclusion}

With a simple constituent model, we predict the masses and 
mean radius of charmonia, lightonia, glueballs and gluelumps. We
We also predict the decay widths of lightonia and glueballs.
The ingredients are constituent quarks and gluons,
and the fundamental SU(3) chromo-electric flux tube. 

We compute the spectrum and wave-functions of
hadrons with angular excitations and maximal $J$.
These hadrons with no radial excitations may be easier 
to identify in the lattice and in the experiments. 
Because we construct all the hadrons with the same confining 
strings, we are able to extrapolate the decay widths of mesons to 
glueballs, in a unified framework. This model can also be
applied to three-gluon glueballs 
\cite{weall3}
and other many-particle systems.

Essentially we expect the angularly excited lightonia and
glueballs to have decay widths of the order of 10\% of the respective
masses. The largest decay width we get, for hadronic masses up to 5 GeV,
are of the order of $\Gamma=0.6$ GeV. Notice that, because we neglected
the direct decay of the constituent gluons, the actual decay widths
of glueballs and hybrids may be larger than the ones predicted here.
We nevertheless point out that the decay width $\Gamma=140$ MeV of 
our $2^{++}$ glueball is identical to the one predicted by Cotanch and 
Williams with vector meson dominance 
\cite{Cotanch:2004py,Cotanch:2005ja}.

However the large mass of the glueballs and of the other studied 
hadrons enable final decay products into a large number of mesons. 
The first decay is expected to result in excited hadrons, and a cascade 
of decays into lower states is expected to produce several pions in the final state. 
Because the partial wave analysis may be quite difficult when, say, more than four 
pions are produced, new efforts in lattice QCD and in the detectors are necessary 
to identify the quantum numbers of the angular excited mesons, glueballs and hybrids.

\acknowledgments
P. B. thanks discussions 
on type-II superconductors with Marco Cardoso and 
Pedro Sacramento, 
on the pomeron with Barbara Clerbaux and Mike Pichowsky, 
on constituent gluons and glueballs with Steve Cotanch, Paola Gianotti,  
Felipe Llanes-Estrada, Orlando Oliveira and Nicoletta Stella
on gluelumps with Alexei Nefediev,
and on excited strings with Fumiko Okiharu.

\onecolumngrid

%
%
\begin{table}[c]
\caption{ \label{ener}
Predicted energies in GeV of charmonia, lightonia,  glueballs and gluelumps as a function 
of $J$. 
We limit our study to maximal $J$ states (no radial excitations and $J=L+S_1+S_2$)
and to energies up to 5 GeV.
}
\begin{ruledtabular}
\begin{tabular}{|l|cccccccccccccccc|} 
hadron $\setminus \ J$ & 0 & 1 & 2 & 3 & 4 & 5 & 6 & 7 & 8 & 9 & 10 & 11 & 12 & 13 & 14 & 15 \\
 \hline
 Charmonium  & & 3.12 & 3.53 & 3.87 & 4.1 & 4.46 & 4.74 & 
4.88 & 5.10 & - & - & - & - & - & - & - 
\\
 lightonium  & & 0.77 & 1.27 & 1.68 & 2.05 & 2.40 & 2.72 & 3.04 
& 3.34 & 3.63 & 3.91 & 4.18 & 4.45 & 4.70 & 4.96 & 5.21 
\\
 Glueball    &&& 2.08 &  & 3.54 &  & 4.67 & & 5.68 & & - & & - & & - & 
\\
 Gluelump    &&& 4.12 & 4.88 & 5.47 & - & - & - & - & - & - & - & - & - & - & - 
\end{tabular}
\end{ruledtabular}
\end{table}
%
%
\begin{table}[c]
\caption{ \label{radi}
Predicted $RMS=\sqrt{\langle r^2 \rangle }$ in GeV$^{-1} \simeq $ 0.20 fm of charmonia, lightonia,  glueballs and gluelumps 
as a function of $J$.
}
\begin{ruledtabular}
\begin{tabular}{|l|cccccccccccccccc|} 
hadron $\setminus \ J$ & 0 & 1 & 2 & 3 & 4 & 5 & 6 & 7 & 8 & 9 & 10 & 11 & 12 & 13 & 14 & 15 \\
 \hline
 Charmonium  & & 2.24 & 3.35 & 4.34 & 5.26 & 6.13 & 6.96 & 7.77 
& 8.54 & - & - & - & - & - & - & - 
\\
 lightonium & & 2.73 & 4.06 & 5.22 & 6.28 & 7.27 & 8.21 & 9.11
& 9.96 & 10.79 & 11.59 & 12.37 & 13.11 & 13.83 & 14.55 & 15.24 
\\
 Glueball  &&& 2.10 &  & 4.06 &  & 5.68 & & 7.13 & & - & & - & & - & 
\\
 Gluelump   &&& 1.86 & 2.82 & 3.64 & - & - & - & - & - & - & - & - & - & - & -  
\end{tabular}
\end{ruledtabular}
\end{table}
%
%
\begin{table}[b]
\caption{ \label{decays}
Estimated decay widths in GeV of lightonia and glueballs as a 
function of $J$. Only the string breaking effect is considered. 
The statistical and systematic error bars are detailed in eq. (\ref{regression}).
}
\begin{ruledtabular}
\begin{tabular}{|l|cccccccccccccccc|} 
hadron $\setminus \ J$ & 0 & 1 & 2 & 3 & 4 & 5 & 6 & 7 & 8 & 9 & 10 & 11 & 12 & 13 & 14 & 15 \\
 \hline
 lightonium & & 0.06 & 0.13 & 0.19 & 0.24 & 0.29 & 0.34 & 0.38
& 0.42 &0.46 & 0.50 & 0.54 &0.58 & 0.61 & 0.65 & 0.68
\\
 Glueball && & 0.14 && 0.33 && 0.49 && 0.64 && - & & - & & - &
\end{tabular}
\end{ruledtabular}
\end{table}


\end{document}